\documentclass[]{acmart}

\usepackage{microtype}
\usepackage{tikz}
\usetikzlibrary{shapes.geometric, arrows.meta}

\usepackage{enumitem}
\usepackage{url}
\usepackage{tcolorbox}
\usepackage{enumitem} 

\usepackage{amssymb}  
\usepackage{booktabs}
\usepackage{tabularx}
\usepackage{caption}
\usepackage{soul}
\usepackage{tcolorbox}

\hypersetup{
  colorlinks=true,
  allcolors=blue,
}

\AtBeginDocument{%
  }

\setcopyright{acmlicensed}
\acmDOI{}
\begin{document}

\setlength{\textfloatsep}{10pt plus 1.0pt minus 2.0pt} 
\setlength{\floatsep}{8pt plus 1.0pt minus 2.0pt}    
\setlength{\intextsep}{8pt plus 1.0pt minus 2.0pt}   

\setlength{\floatsep}{5pt plus 2pt minus 2pt} 

\setlength{\textfloatsep}{8pt plus 2pt minus 2pt} 

\setlength{\abovecaptionskip}{5pt}

\newcommand{\dataset}{\texttt{CHIPS}}

\title{What Do We Mean by “Pilot Study”: Early Findings from a Meta-Review of Pilot Study Reporting at CHI}


\author{Belu Ticona}
\orcid{0000-0002-1003-2618}
\affiliation{%
 \institution{George Mason University}
 \country{United States}}
\email{mticonao@gmu.edu}

\author{Amna Liaqat}
\orcid{0000-0002-5170-1945}
\affiliation{%
 \institution{George Mason University}
 \country{United States}}
\email{aliaqat@gmu.edu}

\author{Antonios Anastasopoulos}
\orcid{0000-0002-8544-246X}
\affiliation{%
  \institution{George Mason University}
  \country{United States}
}
\email{antonis@gmu.edu}

\renewcommand{\shortauthors}{Ticona et al.}

\begin{abstract}
  Pilot studies are ubiquitous in CHI research, yet their role, purpose, and reporting practices remain under-specified. Authors frequently reference pilot studies to justify design decisions or methodological choices, but it is often unclear what constitutes a pilot study, how it differs from other forms of exploratory work, or how pilot findings shape final study designs. We present early findings from an ongoing, large-scale meta-review of pilot study reporting in CHI papers published between 2008 and 2025. Analyzing 904 full papers that explicitly reference “pilot study,” we characterize where pilot studies are reported, how much detail is provided, and how their impact on main studies is described. Our preliminary results reveal substantial variability in reporting structure, limited documentation of pilot outcomes, and frequent ambiguity around pilot contributions. We present this work as a poster to surface emerging patterns and invite feedback from the CHI community, which will inform the next phase.
\end{abstract}

\ccsdesc[200]{Human Centered Computing~Human Computer Interaction} 
\ccsdesc[300]{General and Reference ~ Surveys and Overviews}

\keywords{Pilot Study, Meta-Review, CHI, Meta-CHI}

\received{22 January 2026}

\maketitle

\section{Introduction}
Pilot studies (PS) are ubiquitous in HCI research. CHI papers routinely reference “pilot studies,” “pilot tests,” or “preliminary studies” to justify design decisions, verify procedures, or motivate methodological choices. Yet despite their frequency, the role of pilot studies in HCI remains conceptually vague and empirically underexamined. Unlike fields such as medicine, nursing, and education, where pilot and feasibility studies have well-established definitions, guidelines, reporting standards and even a dedicated research journal, the CHI community lacks a shared understanding of what constitutes a pilot study, why they are conducted, and how they should be reported. Many papers reference pilots “in passing,” without details about design, outcomes, or how the pilot informed the main study. This variability suggests a methodological blind spot in our community.



In this late-breaking work, we aim to start a discussion about the role of pilot studies in the CHI community. We present our preliminary results from a meta-review of pilot study reporting by analyzing 904 CHI papers (2008–2025) that reference pilot-related terminology. Our goals are to:
\begin{enumerate}
    
\item Map the variations in how pilot studies are defined, reported, and justified in HCI
\item Characterize reporting practices (how much detail is given)
\end{enumerate}

Ultimately, we aim to surface opportunities for clearer guidelines and recommendations for the HCI community. This initial work highlights a gap in the methodological foundations of HCI and contributes steps toward a theory of pilot studies tailored to human-centered computing.

\vspace{-0.23cm}
\section{Related Work}

\subsection{Definitions and Theories of Pilot Studies in Other Fields}
\label{sec:psmedicine}
In medicine and the social sciences, pilot studies are well-theorized components of research design, which are constantly revised according to the subject of study \cite{PSJournal}. Their foundational literature provides fine grained distinction between terminology, such as feasibility studies, which assess whether a planned study “can or should be done” (e.g., recruitment, materials, logistics) and pilot studies, which are small-scale versions of a future study to refine procedures, instruments, or parameters.The distinction between these concepts, pilot and feasibility study, is formalized in conceptual frameworks \cite{eldridgeDefiningFeasibilityPilot2016, arainWhatPilotFeasibility2010}, reporting checklists \cite{pfleddererUseGuidelinesChecklists2023}, and journal standards \cite{teresiGuidelinesDesigningEvaluating2022,muraliHowWritePilot2024a}. Pilot studies in these fields support methodological rigor by documenting the rationale, procedure, and outcomes of early-stage testing, being continuously reviewed as an academic practice. Clear understanding of these definitions promote reflexivity, researcher training, and iterative methodological refinement. These are all elements that are highly relevant to HCI but rarely articulated, much less formalized into frameworks or guidelines.


\subsection{Meta-Research in HCI}
CHI has recently seen a rise of interest on the HCI self-reflection of their own practices and structures as a research community (e.g. the MetaHCI workshop in CHI 2025 \cite{oppenlaenderMetaHCIFirstWorkshop2025a, kaltenhauserMetaHCIPractisingReflection2026}). For instance, recent work study bibliometric and meta-methodological approaches \cite{oppenlaenderPresentFutureCitation2025}, the "\textit{LLM-ification}" of CHI submissions \cite{pangUnderstandingLLMificationCHI2025}, what the term "accessibility research" means for HCI community \cite{mackWhatWeMean2021}, contribution types, milestone papers and citation patterns \cite{oppenlaenderKeepingScoreQuantitative2025}, research positionality and subjectivity \cite{singhExploringPositionalityHCI2025}, among others. These studies demonstrate community interest in self-assessment and methodological transparency. Our work contributes to this growing meta-research area by examining a foundational, under-theorized element of the research process: the pilot study.

Only a small number of HCI papers theorize pilot studies directly. Recent work on crowdsourcing finds that pilot studies are inconsistently defined and underreported, often appearing as brief methodological justifications rather than structured investigations \cite{oppenlaenderStatePilotStudy2024}. This work highlights similar concerns: variability in terminology, inconsistent reporting, and ambiguity about purpose.
To date, there is no systematic, cross-venue analysis of how the CHI community as a whole employs pilot studies. 

Despite the prevalence of pilot studies in HCI publications, the field lacks shared definitions, reporting standards, or frameworks for evaluating how pilots contribute to research quality. Authors reference pilot studies in inconsistent ways, sometimes as feasibility tests, sometimes as design iterations, and sometimes merely as rhetorical justification. The absence of conceptual clarity makes it difficult for reviewers to evaluate the appropriateness or rigor of pilot work, and for researchers to learn from each other’s methodologies. This study addresses this gap through a systematic meta-review of how pilot studies are reported in CHI papers.

\section{Methods}
\subsection{Data Creation: the \dataset{} Dataset}

To identify papers that conducted a pilot before their main experimentation, we queried the ACM Digital Library using the keyword "pilot study" in the full text, obtaining an initial dataset of 1887 papers. Since we are interested in understanding how pilots contribute to complete experiments, we limited our search to `full articles' only, obtaining 1098 papers (we filter other types of contributions such as abstracts, posters, etc). Previous studies showed that CHI papers use different terminologies to refer to pilots, therefore we performed a sensitive to evaluate the robustness of our data filtering procedure. We expanded it with other relevant keywords, and quantified its impact in our dataset size. 
Particularly, we included `pilot experiment' (+35, +3.18\%), `preliminary study' (+237, +21.58\%), `initial study' (+206, 18.76\%), `pilot test' (+121, +11.02\%), and `feasibility study' (+56, 5.1\%). In this early stage of work, we focus only on papers including 'pilot study', following our initial filtering criteria. Future work plan to expand our analysis with keywords that significantly increase our dataset (such as 'pilot preliminary study', `initial study', or `pilot test'). 

Once we downloaded the dataset, we proceeded to extract only the texts using a script developed upon the \texttt{PyMuPDF} Python library. We used heuristic criteria to extract from the articles the body text and section titles based on font size, style, and text position. 
This ensures that we filter out content in margins, and after the conclusion section. 
We created a validation set consisting of 10 papers randomly sampled to capture variations over time in the publication style. We iteratively refined our implementation until it consistently extracted the full body text of the validation set, capturing stylistic nuances across the dataset.\footnote{Our pipeline prioritized the identification of main section headings and the preservation of body text. This approach occasionally resulted in noisy text segments where figure and table captions were interleaved with the primary narrative but we deem such minimal noise as acceptable}. 
We also computed tokens per document to filter those with unexpected paper length (<3k), obtaining a cleaned final corpus of 904 documents (see token length distribution in Appendix \ref{sec:appendixData}). We refer to our final corpus as \dataset{} (\textit{CHI Pilot Studies}). 

\subsection{Codebook and Analysis}

We developed our codebook in two stages: 1) Manual Exploration, and 2) LLM-based Annotations. 
In the preliminary stage, a small sample was used to refine the prompt design. These prompts then enabled the automated scaling of annotations across the entire \dataset{} dataset. Then, we focused our analysis on three patterns of reporting structure, for which we conducted human validation.

\subsubsection{Manual Exploration}
We sampled five papers from our dataset to be manually and independently coded by two annotators: one with expertise conducting HCI research and publishing at CHI, and a newcomer in the field with more experience in NLP. This stage of analysis identified an initial set of questions regarding the motivation to conduct a pilot study (e.g. explicit vs implicit purpose), how they are reported (e.g. embedded vs dedicated reporting section, content length, number of participants), and their contribution to the main experiment and future work (e.g. replicability, findings). We expanded our manual exploration with a set of 10 additional articles, manually chosen to represent the general reporting ways identified by \citet{oppenlaenderStatePilotStudy2024}: main-study, detailed-reported, and in-passing.\footnote{This categorization was proposed only for CHI and CSCW papers conducting pilot studies in crowdsourcing.} In the search, we identified a minor proportion of articles conducting a detailed pilot study \textit{before} the main experimentation, explicitly mentioning their results, and why and how they contribute to the main study. Ideally, a recommended research practice would be to report these for research reproducibility and reliability, as done in other fields (see Related Work \ref{sec:psmedicine}).   
Even within this limited subset, we found diverse reporting strategies (e.g. a set of paragraphs without any header; a separate section with `pilot' as title and at the same level of relevance as the main study section; as part of the appendix, etc). 
We also identified a large proportion of in-passing articles that consistently neglect pilot description, and for those the codebook questions would not apply since no enough information is available in the paper. 


\subsubsection{LLM-based Annotation}

Based on this first stage, we decided to consider two ways to structure our codebook questions: as classification tasks, and as open-ended questions. In this stage, we iteratively refined our prompt based on a qualitative performance validation on a set of 15 papers, which we constructed combining all the articles manually analyzed in the previous stage. We explored different ways to specify and contextualize the task and role of the model as annotator, giving concrete examples of phrases to look for (such as `based on pilot feedback, we...', `to validate.., we conducted a pilot...'), and common terminologies. For papers with limited description of the pilot study, the answers obtained in the open-ended questions showed more nuances, occasionally based on the main experimentation methodology rather than responding that not enough information was found. On the other hand, questions framed as classification tasks were able to capture the differences in the reporting structures at different granularity (e.g. pilot and user study as main sections, pilot as a subsection of the methodology description, pilot and experiment described together). However, sometimes the model provided contradictory answers, partially capturing reporting variations in only some questions. In this early work, since we aim to use scaled LLM-based annotations to capture and study reporting patterns and the impact of pilot studies on the main experimentation, we use this methodology to identify concrete examples for each pattern and understand the limitations of the procedure. We used \texttt{gpt-4o-mini} as an LLM annotator, queried through the official Open AI API.\footnote{\url{https://openai.com/api/}} The final version of the prompt can be found in the Appendix \ref{sec:prompt}.  


\subsubsection{Human Validation} We focused the human validation of the LLM-based annotations on papers exhibiting the following reporting patterns: \textit{Pilot as 1st Header}, \textit{PS as Embedded Sub-section}, and \textit{PS Embedded in Method}. We identified the papers for each group based on the classification labels generated by the model (see Appendix \ref{sec:repStructuresDefinition}). A single annotator manually verified the answers, covering at least 10\% of papers for each group.

\section{Findings}

Our results show that our LLM-based methodology facilitates the study of the diverse manners to report pilot studies in CHI papers.

\subsection{Pilot studies are common but rarely treated as distinct studies}

\begin{figure}[h]
\centering
\includegraphics[trim={0cm 0.47cm 0 0.35cm}, clip, width=0.65\textwidth]{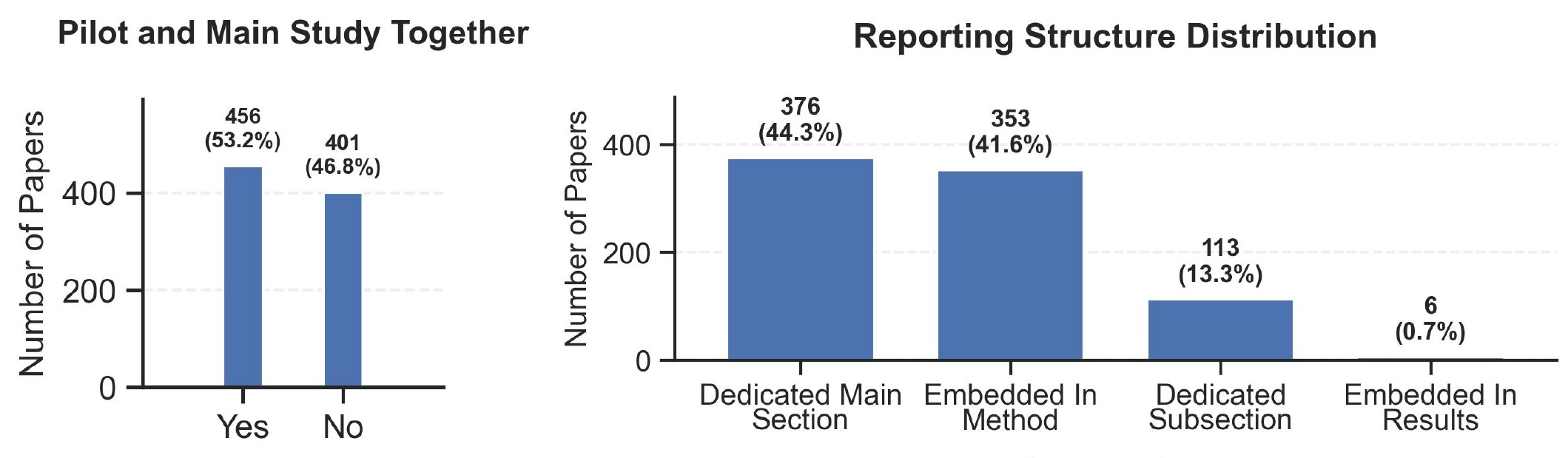}
\caption{Pilot studies reporting practices based on LLM annotations. Frequency of papers that present the pilot and main study described in the same section together (left). Reporting structure distribution in detail: most authors present it as a 'Dedicated Main Section', or `Embedded in Method' (in-passing).}
\label{fig:structureCombined}
\end{figure}

Although pilot studies are frequently mentioned in CHI papers, they are most often not treated as distinct methodological phases. In our corpus of 904 CHI papers referencing “pilot study,” only 37\% report the pilot and main study in clearly separate sections, while 63\% embed pilot descriptions within the main study narrative (see Figure \ref{fig:structureCombined}). When pilots are structurally reported, they most commonly appear either as a dedicated section (44.3\%) or embedded within the methods section (41.6\%). Very few papers report pilot studies in results or discussion sections (<1\%). This structural variability suggests that pilot studies are widely acknowledged but inconsistently framed, often limiting their visibility as independent contributors to research design decisions.

\subsection{Pilot results are usually summarized, not documented for replicability}
\begin{figure}[h]
\centering
\includegraphics[trim={0cm 1cm 0 0.2cm}, clip, width=0.4\textwidth]{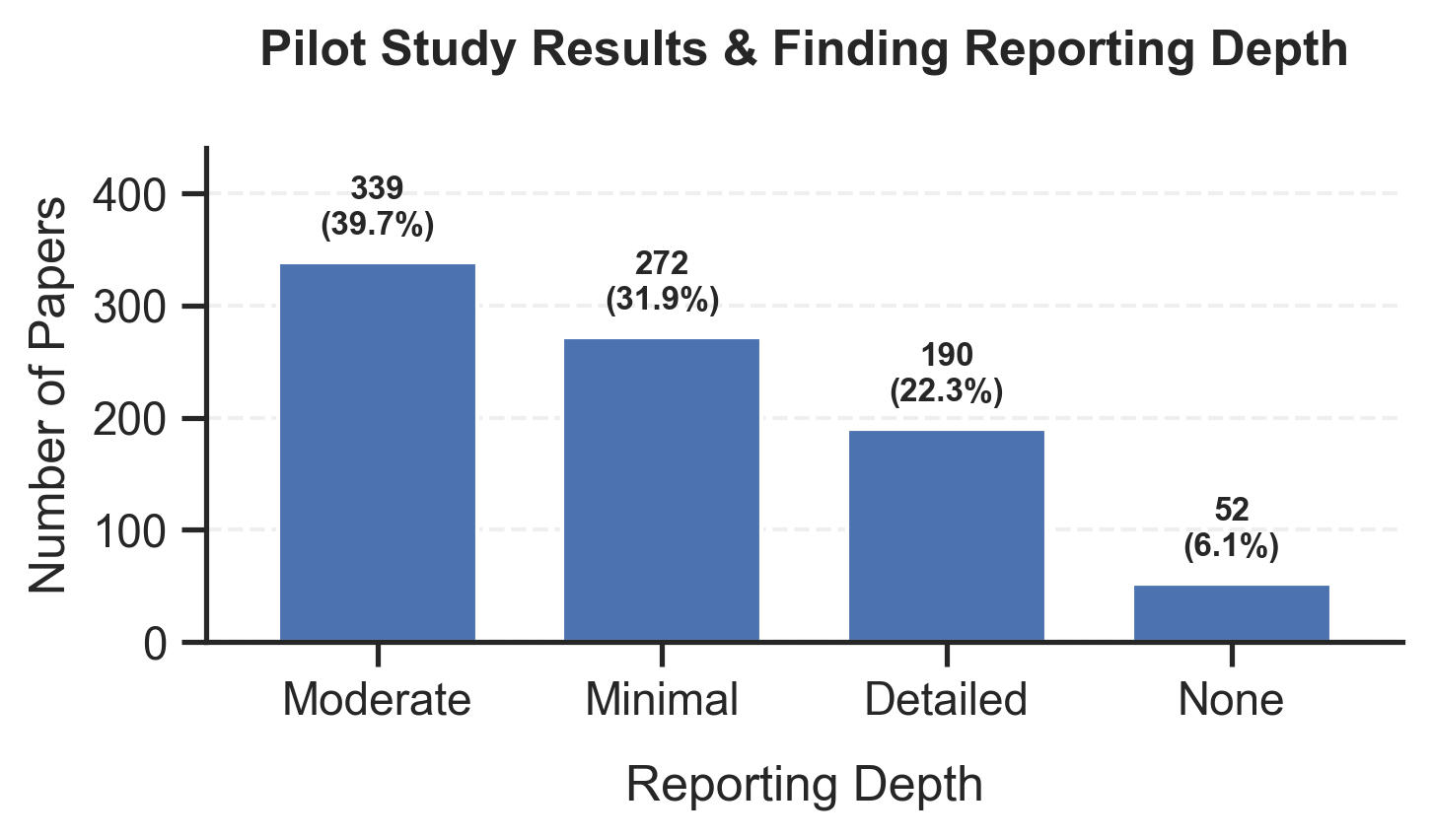}
\caption{Depth of Findings and Result Reporting. Most papers provide 'Moderate' (e.g. a summary, main insights or takeaways without details) or 'Minimal' (e.g. brief mention without details or supporting data) descriptions.}
\label{fig:reportDepth}
\end{figure}

Across the dataset, pilot study results are most often reported at a high level. Only 22.3\% of papers provide detailed pilot findings (e.g., data, statistics, or rich qualitative observations), while 39.7\% offer moderate summaries and 31.9\% include only minimal outcome descriptions. In 6.1\% of cases, papers state that a pilot was conducted without reporting any results at all (see Figure \ref{fig:reportDepth}). This pattern suggests that even when pilot studies shape research decisions, the evidence underlying those decisions is rarely documented in depth. As a result, pilot studies in CHI often contribute to internal decision-making but provide limited methodological insight or replicability value for the broader community.

\vspace{-0.5cm}
\subsection{Pilot studies frequently influence study design, but impacts are under-specified}

\begin{figure}[h]
\centering
\includegraphics[trim={0cm 0cm 0 0.02cm}, clip, width=0.5\textwidth]{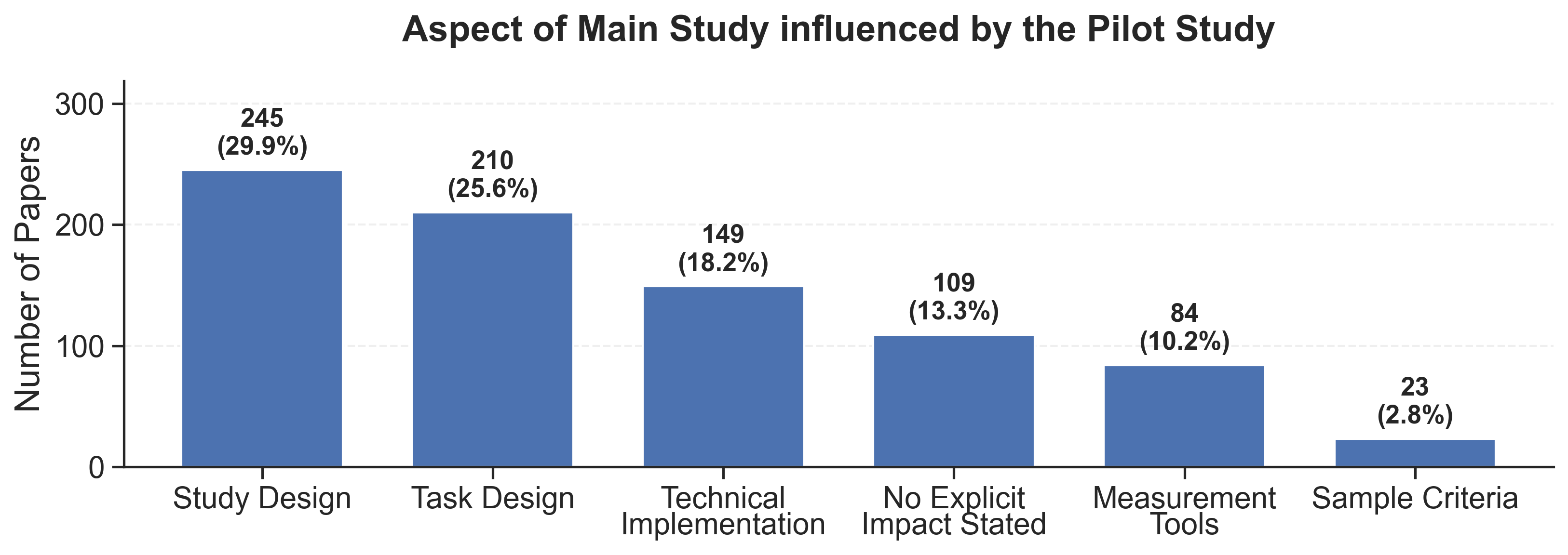}
\caption{Impact of the Pilot Study on Main Study. Task Design covers modification on Activities, Technical Implementation on system/prototype modifications, and Study Design embraces general design changes.}
\label{fig:impactAreas}
\end{figure}

When authors explicitly describe the impact of pilot studies, pilots most commonly influence study design (28.9\%), task design (24.7\%), and technical implementation (17.6\%). Less frequently, pilots inform measurement tools, sampling criteria, study duration, or research questions (see Figure \ref{fig:impactAreas}).

However, many papers reference conducting a pilot study without clearly stating what aspects of the main study were changed as a result. In these cases, pilots function more as implicit justifications than as documented sources of methodological refinement. This lack of explicit linkage between pilot findings and study changes makes it difficult to assess the rigor, necessity, or value of pilot work.

\subsection{Reliability Analysis}

Our results indicate that LLM-based annotations perform best when the pilot description is embedded within the Methods section (achieving >72\% accuracy). When a paper includes a formal methodology section (or synonymous headings such as 'Experimental Design' or 'Experimental Methodology'), this approach effectively identifies 'in-passing' patterns. Conversely, we found that the absence of a well-defined methodology section is a primary cause of failure. Contrary to our expectations, identifying papers where 'Pilot Study' serves as a primary header proved the most challenging, yielding the lowest accuracy (<55\%). Analysis of these misclassifications reveals that the system struggles with section hierarchies, mentions of supplemental material, and instances where the main study itself is framed entirely as a pilot. Finally, for the 'Pilot-as-Subsection' group (~61\% accuracy), we found that inconsistent font styles and formatting often led to misclassifications, particularly when a formal header was absent. 

In general, we found our LLM-based methodology valuable for gaining a detailed understanding of the diverse patterns the CHI community uses to report pilot studies. These findings highlight the potential of LLMs while underscoring the need for further refinement in validation processes. Expert involvement remains crucial to defining the scope and limitations of this study. Future work could explore more sophisticated approaches, such as combining multiple LLMs as annotators or implementing a hybrid human-LLM workflow, among others.

\begin{table*}[h]
\centering
\small
\begin{tabularx}{\textwidth}{@{} >{\bfseries}l c p{1.4cm} p{1.4cm} X @{}} 
\toprule
Group & Acc. & Correct & Incorrect & Primary Failure Reasons \\
Name & (\%) & Examples & Examples & \\
\midrule
Pilot as & 53.1 & \cite{delucaDoesMoodyBoardMake2011} \cite{geHowCultureShapes2024} & \cite{huangMobileMusicTouch2010} \cite{saffoRemoteCollaborativeVirtual2021} & Classified as Header but actually a \textit{Subsection} or found in \textit{Supp. Materials}. \\
1st Header & & & & \\
\addlinespace
PS-as- & 61.5 & \cite{yangHowCanDeep2023} \cite{lawsonValidatingMobilePhone2013} & \cite{schwartzCultivatingEnergyLiteracy2013} \cite{tuModeSwitchingTechniques2012} & Pilot used bolding/formatting but lacked a formal \textit{Sub-header}. \\
Subsection & & & & \\
\addlinespace
PS-embedded & 72.7 & \cite{markCostInterruptedWork2008} \cite{srikulwongComparativeStudyTactile2011} & \cite{linDoesDomainHighlighting2011} \cite{kaziSandCanvasMultitouchArt2011} & Paper lacked a standard \textit{Method} section or pilot was a \textit{Prototype Eval}. \\
(in-passing) & & & & \\
\bottomrule
\end{tabularx}
\caption{Human Validation on LLM-based Annotations (Summary): Comparison of accuracy across reporting structure groups with primary reasons for classification discrepancies. See full Table in Appendix \ref{tab:validation_results}}
\label{tab:compact_validation}
\end{table*}

\vspace{-0.5cm}
\section{Implications and Next Steps}


These findings represent early results from an ongoing, larger meta-review of how pilot studies are defined, reported, and operationalized in CHI research. Our analysis highlights a tension at the heart of current practice: pilot studies are widely used and often shape core methodological decisions, yet they are inconsistently framed, minimally documented, and rarely treated as first-class components of the research process.

This inconsistency has several implications for the CHI community. First, when pilot studies are embedded, summarized, or referenced without clear outcomes, it becomes difficult for reviewers to assess their rigor or necessity, and for readers to understand how early-stage evidence informed final study designs.  Second, the lack of shared expectations around pilot reporting limits opportunities for methodological learning, particularly for early-career researchers who rely on published work to understand norms around exploratory and preparatory research. Finally, ambiguous pilot practices complicate cross-paper comparison and meta-research efforts, obscuring how design and methodological knowledge accumulates over time in HCI.

By surfacing these patterns at an early stage, this poster aims to prompt reflection and discussion within the CHI community about the role pilot studies play in human-centered computing research. We intentionally present this work in the poster venue to gather feedback on which dimensions of pilot practice matter most, where greater clarity would be most valuable, and how future analyses and guidance might better reflect the diversity of CHI methodologies. Community input will directly inform the next phase of our meta-review, including expanded dataset coverage, refined coding dimensions, improved annotation validation, and the development of community-informed recommendations for pilot study reporting.




\bibliographystyle{ACM-Reference-Format}

\bibliography{1-bibliography}

\newpage
\appendix

\section{Data Processing}
\label{sec:appendixData}

\subsection{Token Length Distribution}
We used the \texttt{tiktoken} Python library to calculate the tokens per document using the \texttt{gpt-4o-mini} tokenizer. 

\begin{figure}[h]
\centering

\includegraphics[trim={0cm 0cm 0 0cm}, clip, width=0.45\textwidth]{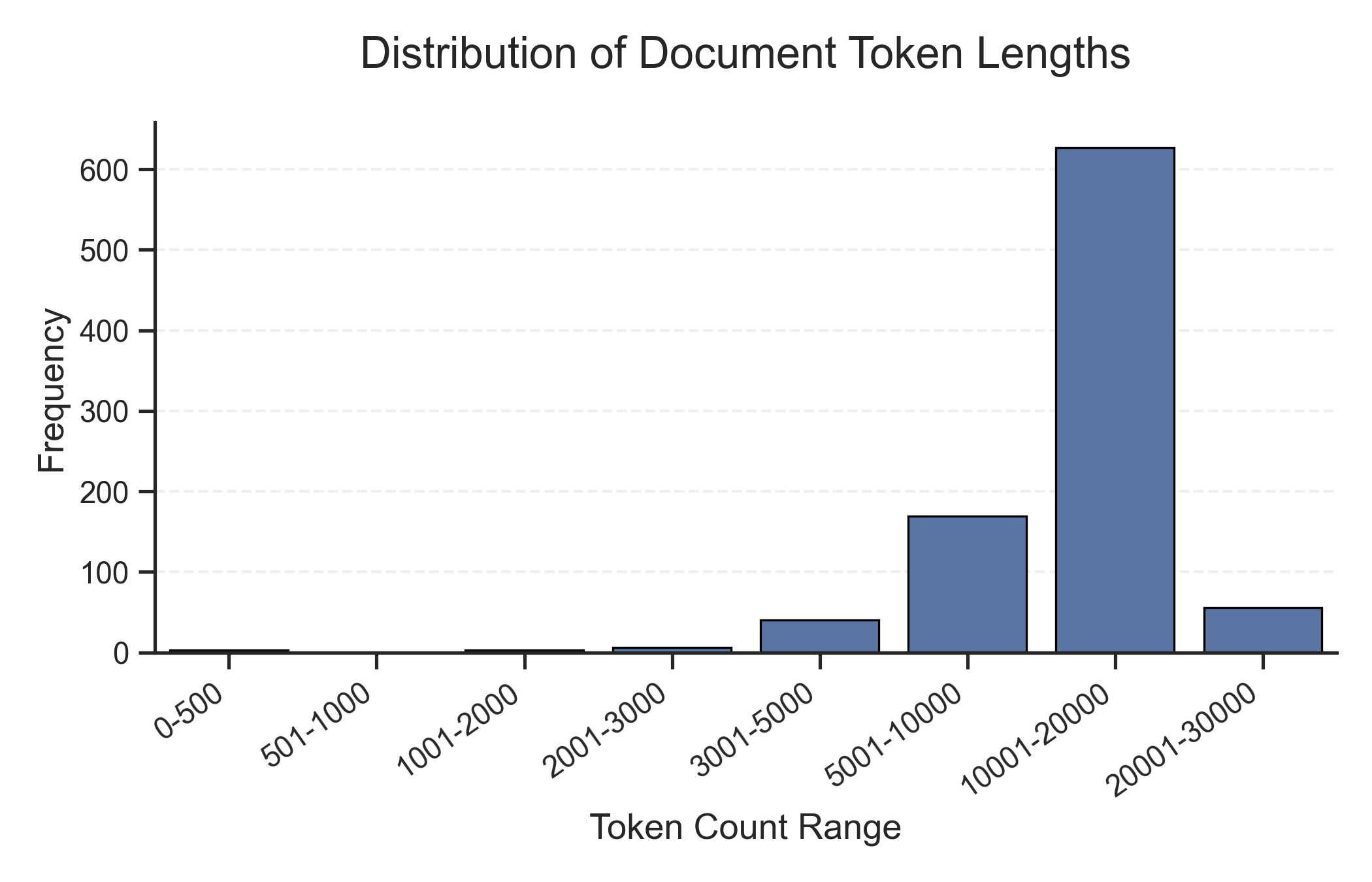}
\caption{Document token length distribution. Token counts are computed using the \texttt{gpt-4o-mini} tokenizer.}
\label{fig:docTokenLengthDist}
\end{figure}

\newpage
\subsection{Annotation Prompt}
\label{sec:prompt}
\begin{tcolorbox}
    \footnotesize
    \begin{verbatim}
system: |
  You are an expert research assistant conducting a meta-study on research practices in HCI. Your goal is to identify 
  generalizable patterns in how pilot studies are reported and contribute to main studies in CHI papers.
  
instruction: |
  Carefully read the provided paper texts and answer the questions below.
  
  GUIDELINES:
  - Base answers ONLY on explicit content in the paper
  - Abstract paper-specific details - focus on research practice patterns, not specific technologies/domains
  - Section headers are typically UPPERCASED and marked with '##'
  - Look for temporal indicators (e.g., "we first conducted...", "prior to the main study...")
  - Check if both pilot AND main study results are reported in the same paper
  - If information is ambiguous or contradictory, mark as UNCLEAR

  Return ONLY valid JSON (no markdown formatting, no code blocks, no extra text).
  
output_format: |
  Return valid JSON in this format:
  {
    "question_id": {
      "answer": "YES | NO | UNCLEAR",
      "evidence": "Brief quote or description (max 20 words)",
      "confidence": "HIGH | MEDIUM | LOW"
    }
  }
  
questions:
  is_ps_preliminary:
    text: "Does this paper report conducting a pilot study as PRELIMINARY work before a separate main study (not as the 
    primary/only study)?"
    clarification: "Look for evidence that both pilot and main study are described in the same paper, with the pilot occurring 
    first temporally."

  ps_has_dedicated_section_only:
    text: "Does the paper report the pilot study and main study in SEPARATE, distinct sections with different headers?"
    answer_options:
      - "YES - Pilot has its own section with a title refering the pilot (e.g., '## PILOT STUDY') AND main study has a 
      different section (e.g., '## MAIN STUDY', '## EXPERIMENT')"
      - "NO - Pilot and main study are reported together in the same section, OR pilot is embedded within main study 
      methodology"
      - "UNCLEAR - Section structure is ambiguous"
      clarification: "Look for section headers marked with '##'. Both studies must have their own distinct sections with 
      clear header. If the pilot is described in a subsection of the methodology or results section that also covers the main
      study, answer NO."

  ps_has_dedicated_space:
    text: "Does the pilot study have its own dedicated area, section or subsection with a clear header title
    (e.g., 'PILOT STUDY', 'PRELIMINARY STUDY', 'PRE-STUDY')?"
    clarification: "Check for explicit section breaks using the symbol '##', not just paragraphs mentioning pilot work."
}
\end{verbatim}
\end{tcolorbox}

\begin{tcolorbox}
    \footnotesize
    \begin{verbatim}


  ps_main_study_same_section:
    text: "Does the paper describe the pilot study and the main studies together in the same section?"
    clarification: "Check for explicit section breaks using the symbol '##', and if the pilot is briefly explained within the
    methodology of the main study."

  ps_reporting_structure:
    text: "How is the pilot study reported in the paper structure?"
    answer_options:
      - "DEDICATED_MAIN_SECTION - Has its own section with a title related to the pilot and where the main study is NOT
      described (e.g., '## PILOT STUDY')"
      - "DEDICATED_SUBSECTION - Has its own subsection within a larger section (e.g., '### Pilot Study')"
      - "EMBEDDED_IN_METHOD - Described within the main methodology section in passing"
      - "EMBEDDED_IN_RESULTS - Mentioned primarily in results/discussion in passing"
      - "UNCLEAR"
    clarification: "Choose a category. Check for section headers (##) and where substantive pilot content appears."

  ps_stated_purpose_categories:
    text: "What reason(s) do the authors explicitly state for conducting the pilot study?"
    answer_options:
      - "VALIDATE_INSTRUMENTS - Test measures/questionnaires"
      - "REFINE_PROTOCOL - Improve procedures/tasks"
      - "TEST_FEASIBILITY - Check if study is viable"
      - "ESTIMATE_PARAMETERS - Determine effect sizes, sample needs"
      - "IDENTIFY_ISSUES - Find problems/confounds"
      - "TRAIN_RESEARCHERS - Practice procedures"
      - "REFINE_TECHNOLOGY - Debug system/prototype"
      - "EXPLORE_PHENOMENON - Initial investigation"
      - "NOT_EXPLICITLY_STATED"
    clarification: "Choose the category that fits the best. Look for phrases like 'to test...', 
    'to refine...', 'to validate...', 'to identify...'. Provide the actual quote as evidence."


  ps_impact_areas:
    text: "What aspects of the main study were influenced by pilot study findings?"
    answer_options:
      - "STUDY_DESIGN - Overall design changes"
      - "MEASUREMENT_TOOLS - Modified/validated instruments"
      - "TASK_DESIGN - Changed tasks or activities"
      - "PARTICIPANT_INSTRUCTIONS - Clarified directions"
      - "DURATION_TIMING - Adjusted study length"
      - "SAMPLE_CRITERIA - Changed inclusion/exclusion"
      - "TECHNICAL_IMPLEMENTATION - System/prototype improvements"
      - "ANALYSIS_APPROACH - Changed analytical methods"
      - "RESEARCH_QUESTIONS - Refined or added RQs"
      - "HYPOTHESIS_REFINEMENT - Modified hypotheses"
      - "NO_EXPLICIT_IMPACT_STATED"
    clarification: "Choose the category that fits the best. Look for explicit statements about how pilot findings led 
    to modifications (e.g., 'based on pilot feedback, we revised...', 'pilot results informed...')."




    \end{verbatim}
\end{tcolorbox}

\begin{tcolorbox}
    \footnotesize
    \begin{verbatim}
 ps_results_reporting_depth:
    text: "How thoroughly are the pilot study findings and results reported in the paper?"
    answer_options:
      - "DETAILED - Comprehensive reporting including specific data (quantitative results, statistics, effect sizes) 
      OR substantial qualitative findings (themes, rich quotes, detailed observations)"
    - "MODERATE - Summary-level findings reported; key takeaways, main insights, or aggregate results mentioned without 
    full detail"
      - "MINIMAL - Brief mention of outcomes or conclusions (e.g., 'pilot showed the task was too long') without supporting 
      data or detailed findings"
      - "NONE - Only states pilot was conducted with no results, outcomes, or findings reported"
    clarification: "Focus on pilot study results only, not main study. Look for: quantitative data (numbers, statistics, 
    graphs), qualitative data (quotes, themes, observations), or explicit findings. If only procedural changes are 
    mentioned without explaining what pilot findings led to them, consider this MINIMAL.
    \end{verbatim}
\end{tcolorbox}

\section{Reporting Structures Groups for Human Validation}
\label{sec:repStructuresDefinition}

\begin{table}[h]
\centering
\label{tab:paper-groups-reporting}
\small 
\begin{tabular}{@{}lccccc@{}}
\toprule
\textbf{Pattern Name} & \textbf{Prelim.?} & \textbf{Standalone?} & \textbf{Ded. Space?} & \textbf{Shared Sec.?} & \textbf{Structure Type} \\ \midrule
Pilot as 1st Header   & Yes & Yes & Yes & No  & Main Section \\
PS (Sub-section)      & Yes & No  & Yes & Yes & Dedicated Sub-sec. \\
PS (Method)           & Yes & No  & Yes & Yes & Embedded in Method \\
PS (Results)          & Yes & No  & Yes & Yes & Embedded in Results \\ \bottomrule
\end{tabular}
\caption{Paper groups categorized based on the pilot reporting location using the prompt questions as variables.}
\end{table}

\begin{table*}[h]
\centering
\small
\begin{tabularx}{\textwidth}{@{} l c c c X @{} } 
\toprule
\textbf{Group Name} & \textbf{Total} & \textbf{Val. \%} & \textbf{Acc. \%} & \textbf{Classification Examples \& Reasoning} \\
\midrule
\textbf{Pilot as 1st Header} & 316 & 10.1\% & 53.1\% & 
\textbf{Correct:} \cite{delucaDoesMoodyBoardMake2011, mulloni360degPanoramicOverviews2012, geHowCultureShapes2024} (e.g., "Section 3: Pilot Study"). \\
& & & & \textbf{Incorrect Reasons:}
\begin{itemize}[noitemsep, topsep=2pt, leftmargin=1.5em]
    \item \cite{huangMobileMusicTouch2010, hollinworthCursorRelocationTechniques2011}: Logged as \textit{Subsection} instead of Header.
    \item \cite{matthewsMoodEngagingTeenagers2011}: Merged \textit{Pilot-Main-Study} section.
    \item \cite{saffoRemoteCollaborativeVirtual2021}: Found in \textit{Supplemental Materials}.
    \item \cite{wuLightWriteTeachHandwriting2021}: No formal title despite details.
\end{itemize} \\ \addlinespace

\textbf{PS Embedded} & 86 & 15.1\% & 61.5\% & 
\textbf{Correct:} \cite{yangHowCanDeep2023, lawsonValidatingMobilePhone2013} (e.g., "Subsection 3.2: Pilot Studies"). \\
(Sub-section) & & & & \textbf{Incorrect Reasons:}
\begin{itemize}[noitemsep, topsep=2pt, leftmargin=1.5em]
    \item \cite{schwartzCultivatingEnergyLiteracy2013}: Used bolding but no sub-header.
    \item \cite{scaffidiTopedEnablingEnduser2008}: Was a standalone \textit{Main Section}.
    \item \cite{tuModeSwitchingTechniques2012}: Mentioned only \textit{In-Passing}.
\end{itemize} \\ \addlinespace

\textbf{PS Embedded} & 108 & 10.2\% & 72.7\% & 
\textbf{Correct:} \cite{markCostInterruptedWork2008, srikulwongComparativeStudyTactile2011} (Within "Method"). \\
(Method) & & & & \textbf{Incorrect Reasons:}
\begin{itemize}[noitemsep, topsep=2pt, leftmargin=1.5em]
    \item \cite{linDoesDomainHighlighting2011}: No distinct Method section in paper.
    \item \cite{kaziSandCanvasMultitouchArt2011}: Pilot categorized as \textit{Prototype Evaluation}.
\end{itemize} \\
\bottomrule
\end{tabularx}
\caption{Manual Validation Summary showing accuracy metrics [cite: 12, 13] and qualitative reasoning for classification discrepancies across patterns[cite: 16, 19, 21].}
\label{tab:validation_results}
\end{table*}

\end{document}